\documentclass[12pt]{article}
\usepackage{graphicx}
\usepackage{amsmath}
\usepackage{amssymb}
\usepackage{epsfig}
\usepackage{subfigure}
\usepackage{float}
\usepackage{verbatim} 
\usepackage{pstricks}
\usepackage{color}
\usepackage{slashed}
\usepackage{multirow}
\usepackage{pstricks}
\usepackage{color}
\usepackage{booktabs}
\usepackage{epstopdf}
\usepackage{multicol}
\newcommand{\asym}{\sigma^{\mathcal{A}}}
\newcommand{\ratio}{\mathcal{R}}

\newcommand\pubdate{\today}

\def\institute{Institute of High Energy Physics, \\Chinese Academy of Sciences, Beijing 100049, China}
\def\support{\footnote{Work supported by the United States Department of Energy
under Grant Contracts DE-SC0012704, and by IHEP under Contract No.~Y7515540U1.}}

\def\Title#1{\begin{center} {\Large #1 } \end{center}}
\def\Author#1{\begin{center}{ \sc #1} \end{center}}
\def\Address#1{\begin{center}{ \it #1} \end{center}}

\newcommand\pubblock{\rightline{\begin{tabular}{l} \\
         \pubdate  \end{tabular}}}
\newenvironment{Abstract}{\begin{quotation}  }{\end{quotation}}
\newenvironment{Presented}{\begin{quotation} \begin{center} 
             PRESENTED AT\end{center}\bigskip 
      \begin{center}\begin{large}}{\end{large}\end{center} \end{quotation}}
\def\Acknowledgements{\bigskip  \bigskip \begin{center} \begin{large}
             \bf ACKNOWLEDGEMENTS \end{large}\end{center}}

\def\beq{\begin{equation}}
\def\eeq#1{\label{#1}\end{equation}}
\def\eeqn{\end{equation}}

\def\beqa{\begin{eqnarray}}
\def\eeqa#1{\label{#1}\end{eqnarray}}
\def\eeqan{\end{eqnarray}}

\let\bar=\overbar

\def\Dslash{\not{\hbox{\kern-4pt $D$}}}
\def\dslash{\not{\hbox{\kern-2pt $\del$}}}

\def\msb{{\bar{\ssstyle M \kern -1pt S}}}

\begin{document}
\begin{titlepage}
\pubblock

\vfill
\Title{Top Quark Properties}
\vfill
\Author{ Cen Zhang\support}
\Address{\institute}
\vfill
\begin{Abstract}

	Recently, experimental collaborations have reported $\mathcal{O}(10)$ upper limits
	on the signal strength of four-top production at the LHC, which already
	leads to competitive constraints on certain type of top-quark
	operators.  On the other hand, developments in the $b$-jet charge tagging algorithms
	open up new possibilities for the direct measurement of the top-quark
	width.  In this talk, we briefly discuss some recent studies in these
	two directions.

\end{Abstract}
\vfill
\begin{Presented}
$10^{th}$ International Workshop on Top Quark Physics\\
Braga, Portugal, September 17--22, 2017
\end{Presented}
\vfill
\end{titlepage}
\def\thefootnote{\fnsymbol{footnote}}
\setcounter{footnote}{0}

\section{Introduction}

The properties of the top quark are partly determined by a set of quantum
numbers, such as charge, spin, color, baryon and lepton numbers.  Deviations of
these numbers from the Standard Model (SM) are possible, but only
discretely.  For this reason these numbers do not provide a general
picture for the interpretation of the precise measurements of top-quark
properties, in particular given the overall precision level of today's
top measurements.  In contrast, the SM Effective Field Theory (EFT)
\cite{Buchmuller:1985jz} provides a more suitable framework to interpret the
small and continuous deviations from the SM.  For example, a measurement on the
$t\bar t\gamma$ vertex is better understood as a constraint on the possible
dim-6 operators such as
\begin{flalign}
	O_{tB}=\left(\bar Q\sigma^{\mu\nu}t\right)\tilde\phi B_{\mu\nu}  \ \,
\end{flalign}
instead of a charge measurement of the top quark itself.

Apart from the quantum numbers, there are other SM parameters that determine
the top quark properties.  Examples include the mass, the Yukawa coupling, and
the CKM matrix elements, etc.  Among them the decay width, $\Gamma_t$, deserves
some more attention.  Even though the top width is not a fundamental parameter
of the SM, decay widths in general are among the most important properties of
fundamental particles.  The most up-to-date constraints on the top-quark width
from CMS and ATLAS still allow for $\mathcal{O}(1)$ deviations from the SM
prediction \cite{CMS:2016hdd,Aaboud:2017uqq}.  This prevents us from directly
interpreting any cross section measurement as a model-independent constraint on the
top-quark couplings, as the branching ratio is always involved in such a
measurement.

In this talk we discuss two recent ideas in these two aspects, respectively
\cite{Zhang:2017mls,Giardino:2017hva}.  The first one is about applying the
SMEFT in the four-top production process, which turns out to be very
sensitive to four-fermion operators, due to multiple insertion of effective
operators in the process.  The second one involves using the $b$-jet
charge identification as a new method to extract the width of the top quark,
which has the advantage of removing completely systematic errors from the
backgrounds.

\section{Four-top production}

At the LHC 13 TeV, this process has a tiny rate, $\approx9$ fb in the SM, five orders of
magnitude lower than $t\bar t$ production, 832 pb.  This process is however
particularly sensitive to new physics.  This can be due to the direct
production of new resonant states which subsequently decay into tops, or to the
contribution from contact four-top operators, which rises as the energy grows.

At first glance, a comprehensive model-independent study of this process in
the context of the SMEFT does not seem to be promising.  The framework
aims to probe the indirect effects from new physics beyond the direct reach of
the LHC.
These effects are expected to show up as relatively small deviations from the
SM predictions, and therefore most of them can be constrained only by precise
measurements. With a current upper bound of $\approx4.6$ times the SM signal
\cite{Sirunyan:2017uyt}, the four-top process is apparently far from
being precise.

Surprisingly, for a very important class of operators, namely the contact
four-fermion interactions with two top quarks and two light quarks, $qqtt$, the
four-top process with only a $\mathcal{O}(10)$ upper bound is already as
powerful as the $t\bar t$ measurement, one of the best measured top-quark processes,
with a percentage error.  This constraining power comes from the fact that the
cross section depends on up to the fourth power of the operator coefficients,
which scales like $(CE^2/\Lambda^2)^4$, where $E$ is the energy of the process,
and $C/\Lambda^2$ is the coefficient of an $qqtt$ operator.  This is because
the $qqtt$ operators can be inserted twice in a $q\bar q\to t\bar tt\bar t$
diagram.  Given the large energy scale related to this process
($\sim\mathcal{O}(1)$ TeV), and the current limits on the coefficient
$C/\Lambda^2$, the factor $(CE^2/\Lambda^2)^4$ significantly enhances the
sensitivity of the four-top process.

To extract reliable constraints on these operators, one has to justify the
validity of the EFT expansion itself, as naively the contributions from dim-8
and higher operators scale the same way in $1/\Lambda$ as the
$(CE^2/\Lambda^2)^4$ terms.  It is useful to distinguish between two
kinds of ``expansions''.  The EFT expansion comes from integrating out heavy
degrees of freedom at the energy scale $\Lambda_{NP}$ (to be distinguished from
the non-physical $\Lambda$), a procedure whose validity is achieved by
imposing an analysis cut, $M_{cut}<\Lambda_{NP}$.  The related error due
to truncating higher dimensional terms is estimated by
$E^2/\Lambda_{NP}^2<1$.  This is different than the ``expansion'' due
to multiple insertion of dimension-six effective interaction and squaring the
amplitude, where the ``expansion parameter'' is $CE^2/\Lambda^2>1$.
This second ``expansion'' is not related to EFT validity, because
there are no more terms after the fourth power of $CE^2/\Lambda^2$, and so no
truncation happens.  Simply put, the EFT can be valid as far as all terms in a
series of $CE^2/\Lambda^2$ are kept, and $M_{cut}<\Lambda_{NP}$ is imposed.

As an example, it has been shown in Ref.~\cite{Contino:2016jqw} that the dim-6
squared terms could dominate over the dim-8 operator contributions, without
invalidating the EFT expansion.  The situation for the four-top process is
similar.  By assuming that the underlying theory is characterized by one scale
$\Lambda_{NP}$ and one coupling $g_*$, with a reasonable power counting rule,
one can justify the truncation of SMEFT at dim-6.

\begin{figure}[htb]
\centering
\includegraphics[width=.26\linewidth]{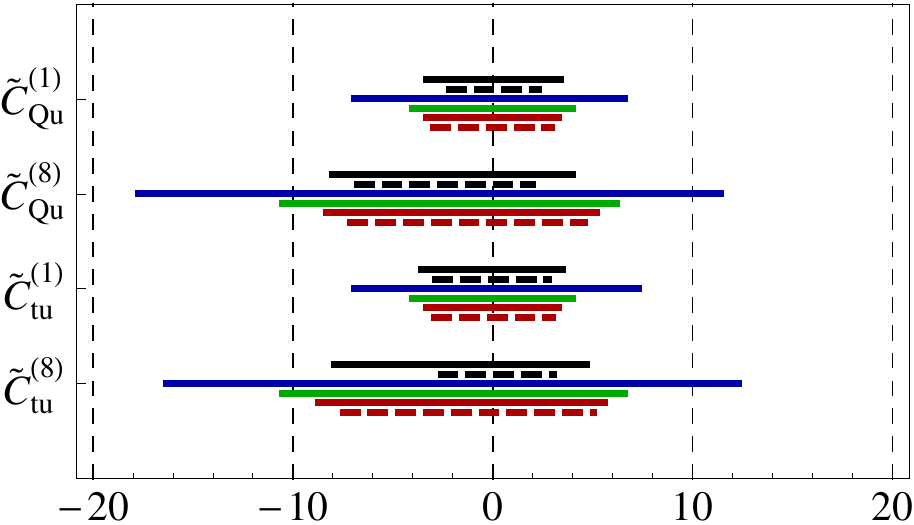}
\includegraphics[width=.26\linewidth]{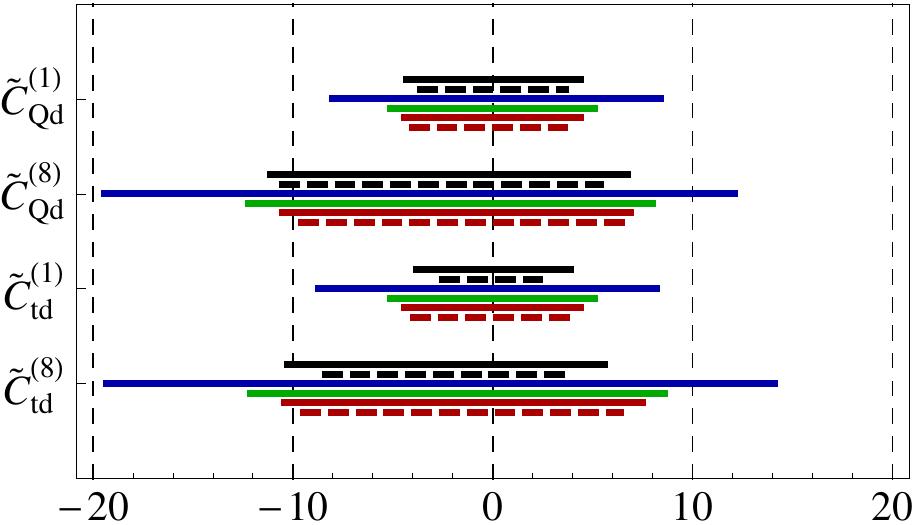}
\includegraphics[width=.45\linewidth]{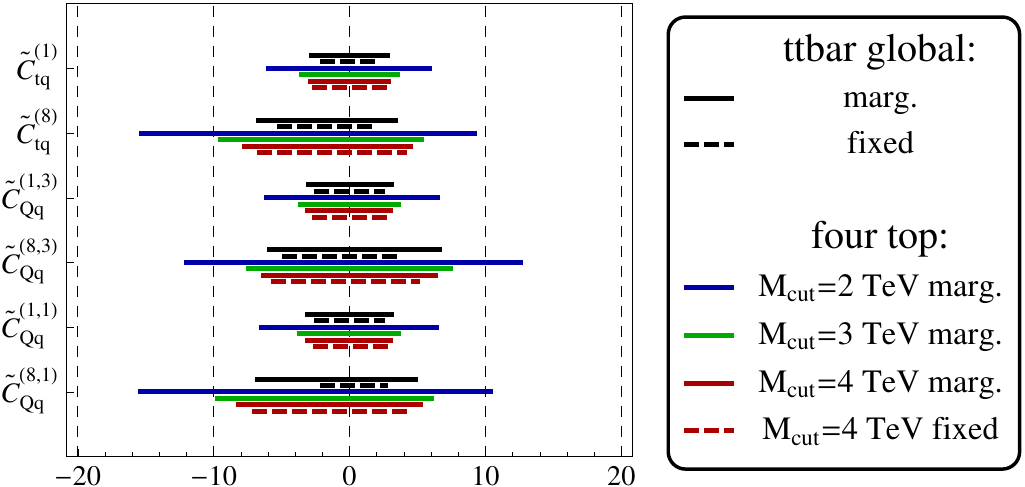}
\caption{Fixed (i.e.~one operator at a time) and fully marginalized
(i.e.~all other operators floated) constraints for all $qqtt$ operators, from
four-top and from $t\bar t$ measurements, at 95\% CL.  The $t\bar t$
constraints are from our global fit, while the four-top constraints are from
the 300 fb$^{-1}$ projection.}
\label{fig:4t}
\end{figure}

With this justification we are ready to present the projected limits from
four-top production.  In Figure~\ref{fig:4t} we show these limits and compare
with the current limits from top-pair production.  The four top constraints
correspond to an integrated luminosity of 300 fb$^{-1}$, while the top-pair
constraints come from a global fit taking into account the major inclusive
measurements so far as well as a differential $m_{t\bar t}$ measurement.  While
the details and the definition of the operators can be found in
Ref.~\cite{Zhang:2017mls}, the comparison presented in Figure~\ref{fig:4t}
clearly shows that, with $M_{cut}=3$, 4 TeV, the four-top process can be as
good as those from a $t\bar t$ global fit.  We want to emphasize that these
constraints are still conservative, due to imposing $M_{cut}$ and assuming the SM
signal shape, and in practice better results can be expected from a tailored
experimental analysis.  We also remind the reader that the cost of such an
enhanced sensitivity is a relatively large value of $M_{cut}$.  In the long
term, however, we believe that in any case new states below this energy scale
are likely to be excluded by explicit resonance searches.

\section{Top-width measurement}

So far, direct measurements of the top-quark width are only through
(partially) reconstructing the top-quark kinematics.  The most recent limits
from CMS \cite{CMS:2016hdd} and ATLAS \cite{Aaboud:2017uqq} still allow for
$\mathcal{O}(1)$ deviation from the SM value.  With an undetermined top-quark
width, physics beyond the SM that enhances the major production mechanisms and
at the same time increases the top-quark width, e.g.~through undetectable decay
channels, can still leave the measured cross sections unchanged, and will not
be directly excluded by existing measurements.  Improving the direct width
measurement is the best way to break this degeneracy.

\begin{figure}[htb]
\centering
\includegraphics[height=1.2in]{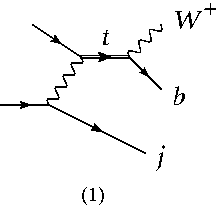}
\includegraphics[height=1.2in]{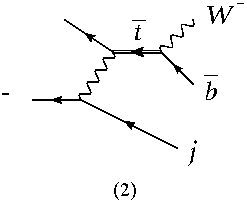}
\includegraphics[height=1.2in]{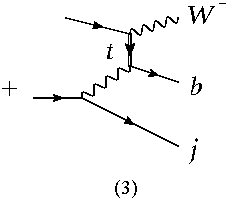}
\includegraphics[height=1.2in]{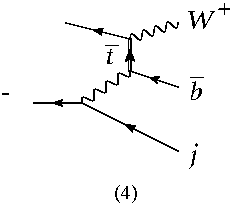}
\caption{A combination of the four $bW$ scattering channels corresponds to
the $b$-jet charge asymmetry.}
\label{fig:bw4}
\end{figure}

It is well known that the decay width can be extracted from the 
ratio between the on- and off-shell cross sections \cite{Caola:2013yja}.
Consider the four $bW\to bW$ scattering processes, which differ in the charges
of the scattering particles, as depicted in Figure~\ref{fig:bw4}.  The $bW^+\to
bW^+$ and $\bar bW^-\to \bar bW^-$ scattering processes are simply the
$t$-channel single top processes, where the top quarks are in the $s$-channel,
from which both on-shell and off-shell cross sections can be measured.
Alternatively, in $bW^-\to bW^-$ and $\bar bW^+\to \bar bW^+$ scattering the
top-quark is in the $t$-channel, which is always off-shell, and could offer
complementary information.  The main difficulty however is that the off-shell
cross sections are hard to measure, due to large backgrounds from QCD, $t\bar t$,
and $tW$ production processes.

The recent developments in the $b$-jet charge tagging techniques,
e.g.~Ref.~\cite{jvc}, open up a new possibility. It turns out that a
specific charge combination of the four $bW$ scattering channels is virtually
free of any background related systematic uncertainties.  This combination is
illustrated in Figure~\ref{fig:bw4}, which is equivalent to the $b$-jet charge
asymmetry (inclusive in $W$ charge), $\asym$, of the process $pp\to Wbj$.  The
on-shell and off-shell ratio of this asymmetry, $\ratio$, is an ideal
observable from which the width could be extracted, for the following reasons:
\begin{itemize}
	\item The four $bW\to bW$ amplitudes are related by crossing symmetry,
		so they represent the same physics but with different
		kinematics.  New physics modifications to the couplings are
		supposed to cancel out when taking the on-/off-shell ratio.
	\item The QCD background vanishes at the LO: 
		$\sigma_{QCD}(bW^\pm j)=\sigma_{QCD}(\bar bW^\pm j)$,
		i.e.~the backgrounds of diagrams (1), (4) cancel out,
		and that of (2), (3) cancel out.
	\item The $t\bar t$ and $tW$ backgrounds vanish at the LO:
		$\sigma_{t\bar t,tW}(bW^\pm j)=\sigma_{t\bar t,tW}(\bar
		bW^\mp j)$, i.e.~the backgrounds of diagrams (1), (2) cancel
		out, and that of (3), (4) cancel out.
	\item The uncertainty of the charge tagging efficiency cancels out
		in the ratio.  This is because the tagged charge asymmetry 
		is given by
			$\asym(\mathrm{tagged})=(2\epsilon-1)\asym$,
		where $\epsilon$ is the charge tagging efficiency.
		This parameter can reach $\approx65\%$ \cite{jvc}, if the
		correct charge tagging rates for $b$ and $\bar b$ are kept to
		be the same.  When taking the on-/off-shell ratio, the
		$(2\epsilon-1)$ factor cancels out, leaving no systematic error
		from the tagging efficiency.
\end{itemize}

The reasons for the cancellations of the backgrounds are explained in
Ref.~\cite{Giardino:2017hva} in detail.  While NLO corrections could in
principle generate some contributions to $\asym$, it can be shown, by explicit
calculations, that they are negligible.  The signal process, on the other hand,
gives rise to a non-vanishing $\asym$ at the LO, due to the charge asymmetry of
the $W$ boson in the proton.  
It can be further shown that the uncertainty on $\ratio$ due to radiative
correction is only about $1.5\%$.  With all this information we can estimate
the projected limits at the LHC.  Note that even though the backgrounds
do not contribute to the systematic uncertainty of $\asym$, they do give rise
to the statistical uncertainties, which are the dominant source of uncertainties
in this approach.  We thus expect this approach to reach a good
precision at high luminosity with enough statistics, at a few hundred MeV.
This is in contrast with the current methods used by the experimental
collaborations, where the systematic effects will eventually dominate, if
not already. The results for the 13 TeV run are presented in
Table~\ref{tab:result}.

\begin{table}[h]
\begin{center}
	\begin{tabular}{cccc}
		\hline\hline
	Luminosity [fb$^{-1}$] & 30 & 300 & 3000
	\\\hline
	Limits [GeV] & [0.40,2.30] & [1.01,1.73] & [1.14,1.60]
	\\\hline\hline
	\end{tabular}
\end{center}
	\caption{\label{tab:result}One-sigma exclusion limit on $\Gamma_t$,
expected at LHC 13 TeV.}
\end{table}

\Acknowledgements
I am grateful to P.~P.~Giardino for the collaboration on the investigation of
probing the top-quark width using $b$-jet charge identification.

\end{document}